# Assessing the Macroeconomic Impacts of Disasters: an Updated Multi-Regional Impact Assessment (MRIA) model

**Surender Raj Vanniya Perumal [a]\*, Mark Thissen [b, c] , Marleen de Ruiter [a], and Elco E. Koks [a]**

[a] *Institute for Environmental Studies (IVM), Vrije University Amsterdam, Amsterdam, The Netherlands;*

[b] *Spatial Economics, Vrije Universiteit Amsterdam, The Netherlands;*

[c] *Netherlands Environmental Assessment Agency (PBL), The Hague, The Netherlands.*

\*Corresponding author: Surender Raj Vanniya Perumal, s.r.vanniya.perumal@vu.nl

**Abstract**

Disasters often impact supply chains, leading to cascading effects across regions. While unaffected regions may attempt to compensate, their ability is constrained by their available production capacity and logistical constraints between regions. This study introduces a Multi-Regional Impact Assessment (MRIA) model to evaluate the regional and macroeconomic consequences of disasters, capturing regional post-disaster trade dynamics and logistical constraints. Our findings emphasize that enhancing production capacity alone is inadequate; regional trade flexibility must also be improved to mitigate disaster impacts. At the regional level, disaster-affected areas experience severe negative impacts, whereas larger, export-oriented regions benefit from increased production activity. Additionally, we propose a sectoral criticality assessment alongside the more common sensitivity and incremental disruption analysis, which effectively identifies sectors with low redundancy while accounting for the potential for regional substitution in a post-disaster scenario.

1.   **Introduction**

Natural hazards have broad-reaching impacts, disrupting the social and economic functioning of societies. They can cause significant damage to infrastructure, capital, and, labour (Rose & Liao, 2005; Thekdi & Santos, 2016; Wei et al., 2022; Xia et al., 2018). As global trade connects regions, the economic consequences of large-scale disasters can extend beyond regional boundaries, affecting economies in other regions. For example, frequent tropical cyclones in Taiwan disrupt the production of microchips, causing ripple effects across the global computer and electronics sector, including major economies such as China, the United States, and Japan (Allianz Research, 2024). Therefore, understanding these cascading economic impacts of disruptions is crucial for effective disaster response and adaptation planning.

Various methods have been used to assess the macroeconomic effects of natural hazards, including statistical (or) econometric analysis, input-output (IO) analysis, and computable general equilibrium (CGE) analysis (Botzen et al., 2019; Kelly, 2015). Econometric analysis uses historical data to understand the impacts of disasters on long-term economic growth using statistical techniques (Balakrishnan et al., 2022; Botzen et al., 2019; Cavallo et al., 2013; Skidmore & Toya, 2002) rather than the economic output. In contrast, IO and CGE models are designed to capture the cascading effects of disasters across sectors, as well as interregional impacts in multi-regional models. IO models, while simpler and linear, tend to provide higher-end estimates of impacts due to their rigid structure, while neo-classical CGE models, which include price and substitution effects, may provide lower-end estimates of impacts, particularly in the immediate aftermath of a disaster since they allow for more flexibility (Rose, 2004). Also, CGE models are more intricate, which makes them difficult to construct and costly to estimate (Oosterhaven & Bouwmeester, 2016) compared to IO models.

In the past decades, IO-based models have gained more attention in assessing disaster consequences (Hallegatte, 2008; Kelly, 2015). Yet, while natural hazards hinder the production capacities of firms (and sectors), such a supply-side shock cannot be handled by a traditional IO model. Conventional IO models, including Inoperability Input-Output Models (IIMs), therefore typically address supply-side disruptions by representing them as equivalent reductions in demand (Anderson et al., 2007; Haimes & Jiang, 2001; Jonkeren & Giannopoulos, 2014; Okuyama, 2004; Rose et al., 1997; Thekdi & Santos, 2016). Also, a traditional IO framework is linear, which does not allow for adaptive flexibilities such as inter-regional substitution, production extension, and inventory capacities. Taking this into account, hybrid models have been developed that retain the basic framework of IO modelling with added relevant flexibilities available in CGE models. For example, the ARIO model (Adaptive Regional Input Output) developed by Hallegatte (2008) can handle the reduced production capacities and demand changes after the disaster simultaneously with flexibilities in production extension. The model is widely used to assess the impacts of disasters (Guan et al., 2020; Hu et al., 2023; Zeng & Guan, 2020) and has been improved to incorporate inventory (Hallegatte, 2014) and multi-regional substitution flexibilities (Liu et al., 2023). Also, agent-based models (ABMs) have been built upon the IO framework where the impacts are estimated based on the behaviour of economic agents e.g., production and consumption sites after the disaster (Bierkandt et al., 2014; Otto et al., 2017; Poledna et al., 2023; Wenz et al., 2014; Willner et al., 2018). With the advancement of computational capabilities, researchers have also explored the capacity of (non-) linear programming approaches combined with the IO framework in disaster impact modelling (Baghersad & Zobel, 2015; Bouwmeester & Oosterhaven, 2017; Oosterhaven & Bouwmeester, 2016). For example, Oosterhaven &

Bouwmeester (2016) developed a novel optimization-based approach to simulate disruption in inter-regional IO tables. This method aims to identify an alternative equilibrium by minimizing the information gain between pre- and post-disaster IO tables.

Expanding on these advancements, Koks & Thissen (2016) developed a Multi-Regional Impact Assessment (MRIA) model, an adaptive linear programming model based on the supply-and-use framework (an extension of the IO framework). The MRIA model improves upon traditional IO models in three key ways. First, it explicitly accounts for supply-side shocks, such as reduced production capacity, and demand shocks via reconstruction demands after the disaster. Second, the model being multi-regional, allows substitution through increased production capacity in non-disaster-struck regions and also allows rationing of products. This approach enables the assessment of not only the losses but also the gains outside the disaster-affected region. Lastly, the use of a supply-use framework introduces the ability to allow for inefficient production to meet post-disaster demands, which is novel and first of its kind.

The focus of this paper is twofold: first, to develop an MRIA model that builds upon the foundations established by Koks & Thissen (2016), while addressing its key limitations. While the model presented by Koks & Thissen (2016) estimates the net import of a product required by a region following a disaster, it did not properly capture the inter-regional redistribution of trade flows, i.e., the value of products exported from one region to another in the aftermath of a disaster. The revised formulation presented in this paper explicitly models the origin and distribution of disaster-induced trade, providing a more accurate representation of post-disaster supply chains. Moreover, the post-disaster trade in Koks & Thissen (2016) is unrestricted, constrained only by the production capacities of sectors in unaffected regions, and regions that were not engaged in trade pre-disaster were unable to participate. However, the ability of an unaffected region to export products to a disaster-struck region depends heavily on the existing logistics system between the regions. The updated model includes this post-disaster logistical constraint, distinguishing it as one of the first disaster impact modelling studies to do so. The logistical constraint between regions is imposed by limiting the post-disaster trade between them, as modelling the actual logistics system (e.g., transportation hubs) is beyond the scope of this paper. Finally, rationing of products results in welfare loss, primarily impacting consumers. In the model by Koks & Thissen (2016), imports are prioritized over rationing of products through weight adjustments in the objective function. However, the updated model employs a multi-step approach to ensure minimal rationing of products.

Secondly, we demonstrate the capabilities of our model in performing a range of analyses relevant to disaster risk applications. We perform three distinct types of analysis: multi-sector disruption analysis (simulating a natural hazard disruption), criticality analysis to prioritise crucial economic sectors, and incremental disruption analysis to identify critical disruption thresholds of an economic sector. Notably, this is one of the first studies to conduct a criticality analysis across an entire economic system. Existing indices for sectoral prioritisation, e.g., Yu et al., (2014), often don't account for the flexibility within the economic system (e.g., production possibility limits) and the extent to which sectors are stressed. For instance, a sector of higher output with regional substitution options available during a disaster may not be as critical as a sector with lower output but that has a unique role in the supply chain with no viable regional substitutes. Our approach to estimating criticality addresses both factors. Also, we focus only on the regions of the Netherlands at the NUTS-2 level as a test case for all our analyses. However, the model can be easily scaled up to a larger set of European regions, sectors, and product groups. The rest of the paper is organized as follows: Section 2 explains the model; Section 3 presents the results of the analyses mentioned above; Section 4 provides a discussion of the results along with recommendations; and Section 5 provides a summary of the study.

## 2. Methods

### 2.1 Supply and Use Dataset

For this study, we utilize the Supply and Use Tables (SUTs) from the recently developed RHOMOLO V4 dataset (García-Rodríguez et al., 2025). This global dataset includes interregional trade flows, representing the supply and use activities, across NUTS-2 (Nomenclature of Territorial Units for Statistics) regions in Europe and key global economies such as the United States, and China. It features a classification of 55 sectors (corresponding to 64 product groups) based on the NACE level 2 framework, with some sectors aggregated. A list of all the sectors is presented in Supplementary Table 1. The dataset builds upon the FIGARO (Full International and Global Accounts for Research in Input-Output Analysis) intercountry tables (Remond-Tiedrez & Rueda-Cantuche, 2019), which are further disaggregated into interregional flows using freight transport, business travel, and flight data (Thissen et al., 2019). For detailed insights into the dataset construction, see García-Rodríguez et al. (2025). In this study, we focus exclusively on the 12 NUTS-2 regions of the Netherlands (refer to Supplementary Figure 1) while aggregating other regions into a single rest of the world category.

### 2.2 MRIA model

MRIA is a Supply and Use multisector multiregional macroeconomic model solved using linear optimization techniques to evaluate the macroeconomic and regional consequences of disasters. The model incorporates adaptive responses by economic agents (in this case, sectors), allowing sectors to increase production up to a maximum production level, ration final consumption when production limits have been reached, and export surplus production to meet the demands of the economy elsewhere. The complete set of equations that define the model is presented in Table 1, and a description of the symbols used is provided in Supplementary Table 2.

We discuss our model, explaining the key assumptions involved. The MRIA is a demand-driven model, where the demand determines overall production i.e., sectors produce to satisfy the demand for products in place e.g., household and government consumption, exports. Furthermore, products will only be rationed if supply constraints in the post-disaster economy prevent meeting the demand. Similar to Koks & Thissen (2016), this model also builds on a supply-and-use framework. Sectors produce a mix of products (presented in the supply table) by using a specific mix of inputs (presented in the use table), which represents their production technologies. The following example illustrates production technology and inefficiencies in the production process. Consider the production technology of the oil refining industry, which primarily uses crude oil, electricity, and water in a specific proportion to produce gasoline (fuel) as its primary product. During refining, additional by-products, such as petrochemical feedstocks, are generated and supplied to industries such as rubber and plastic manufacturing. The mix of main and by-products is assumed as an intricate part of the production technology used and is therefore assumed to remain fixed, both before and after a disaster. Following a disaster, assume that the demand for fuels increases significantly while the demand for petrochemical feedstocks remains unchanged. To meet the increased demand for fuel, the refinery must elevate overall production levels, inevitably producing more petrochemical feedstocks as a by-product due to its available production technology. This additionally produced by-product that is not used in the rest of the economy is the inefficiency (see cost discussion later in Section 2.3) we refer to. MRIA allows inefficient production to happen post-disaster, which is reflected in its constraint where the supply of a product should be greater than or equal to its demand (see Table 1; Model constraints).

**Table 1: MRIA model equations**

|  | Step 1: Minimise rationing | Step 2: Minimise production and disaster trade | Step 3: Estimate the production equivalent of rationing |
|---|---|---|---|
| Objective | $z_1 = min\left(\sum_r \sum_p v_{r,p}\right)$ | $z_2 = min\left(\sum_r \sum_s x_{r,s} + \alpha \sum_{r'} \sum_r \sum_p t_{r',r,p}\right)$ | $z_3 = min\left(\sum_r \sum_s x'_{r,s}\right)$ |
| Supply and demand | $s_{r,p} = \sum_s C_{r,s,p} x_{r,s} + \sum_{r'} t_{r',r,p}$ $d_{r,p} = \sum_{r'} \sum_s B_{r,p,r',s} x_{r',s} + \overline{f}_{r,p} + \overline{e}_{r,p} + \overline{n}_{r,p} - v_{r,p} + \sum_{r'} t_{r,r',p}$ | $s_{r,p} = \sum_s C_{r,s,p} x_{r,s} + \sum_{r'} t_{r',r,p}$ $d_{r,p} = \sum_{r'} \sum_s B_{r,p,r',s} x_{r',s} + \overline{f}_{r,p} + \overline{e}_{r,p} + \overline{n}_{r,p} - v_{r,p} + \sum_{r'} t_{r,r',p}$ | $s_{r,p} = \sum_s C_{r,s,p} x'_{r,s}$ $d_{r,p} = \sum_{r'} \sum_s B_{r,p,r',s} x'_{r',s} + \overline{\overline{v}}_{r,p}$ |
| Model constraints | $s_{r,p} \geq d_{r,p}$ $x_{r,s} \leq \overline{x}_{r,s} \delta_{r,s}$ $t_{r',r,p} \leq \left(\sum_s B_{r',p,r,s} \overline{x}_{r,s}\right) \phi_{r',r,p}$ $v_{r,p} \leq \overline{f}_{r,p} + \overline{e}_{r,p} + \overline{n}_{r,p}$ | $s_{r,p} \geq d_{r,p}$ $x_{r,s} \leq \overline{x}_{r,s} \delta_{r,s}$ $t_{r',r,p} \leq \left(\sum_s B_{r',p,r,s} \overline{x}_{r,s}\right) \phi_{r',r,p}$ $v_{r,p} \leq \overline{v}_{r,p}$ | $s_{r,p} \geq d_{r,p}$ |

The possibility of rationing is modelled using a two-step procedure. Firstly, under the given supply constraints (e.g., production extension limits), the model determines the minimum possible rationing (see Objective Function; Step 1; Table 1). This minimized rationing level is then used as an upper limit for rationing in Step 2, where we assume that sectors from all regions minimize their production to satisfy product demand (after minimal rationing). Hence, given the amount of rationing, we search for the least cost solution that satisfies the remaining demand. This is implemented by the second objective function, minimizing production and post-disaster trade (see Objective Function; Step 2; Table 1). This objective function incorporates a weighing parameter ($\alpha$) that represents logistical costs related to trade. We calibrate $\alpha$ to the minimum value of 1.25 such that the model still reproduces the existing trade flows and regional production. In this way, the model generates no additional trade or changes in production in the absence of a disruption (i.e., it is replicating the baseline scenario).[1] Supplementary Figure 2 presents the outcomes of this calibration procedure. Further, the model estimates the equivalent production $(x'_{r,s})$ required if the rationed products were to be produced within the economy. This is accomplished by incorporating the rationed products as final demands in the model (see Step 3, Table 1). The implication of this step is presented in the cost discussion later (see Section 2.3).

---

[1] When $\alpha$ equals 0, the model does not revert to the baseline solution; instead, it returns to an alternate minimum cost equilibrium involving additional trade and a change in regional production.

Further, the model's flexibility is governed by two key parameters: the production extension factor ($\delta_{r,s}$) and the trade flexibility factor ($\phi_{r',r,p}$). The production extension factor represents a sector's capacity to increase production following a disaster. This factor is also used as a channel to simulate disruptions, where a value less than 1 indicates reduced production capacity due to the disaster, and a value greater than 1 reflects the potential for increasing the production. The pre-disaster production ($\bar{x}_{r,s}$), adjusted by this factor, is applied as a constraint on the post-disaster production capacity of each sector (see Model Constraints; Table 1).

The trade flexibility factor, one of the main contributions of this paper, represents the proportion of additional trade permitted between regions for a specific product after a disaster, relative to its initial trade level. This factor accounts for the trade constraints that arise following a disaster due to the logistical systems in place. We can best explain this using an example: Suppose region A imports food products from Region B. The food products from region B are stored in a warehouse with capacity 'q' before being moved to region A. After a flood event in Region A, its local food production declines, increasing the need for imports from Region B. However, Region B cannot rapidly scale up its logistics system, particularly in the short term, and can only arrange for a limited extension of its storage capacity, say 'Δq'. The ratio of this additional capacity (Δq) to the initial capacity (q) defines the trade flexibility factor, which is multiplied by the initial trade to set the upper bounds for disaster trade between the regions (see Model Constraints; Table 1). Thus, including trade rigidities improves the model by capturing the logistical costs of changing the existing trade network. Moreover, this approach prioritizes regions seeking additional imports from their conventional suppliers, as these suppliers are likely to have the highest disaster trade capacity. It also ensures that regions not previously involved in trade will remain the same, as establishing new trade links in the short term after a disaster is unlikely. More importantly, this approach gives the modeler greater control over how regions interact after a disaster. For example, supply and use tables (or input-output tables) show the aggregated values of products supplied and used across regions. Region A may be a major producer of wheat, while Region B is a major producer of rice, but both are recorded under the same agricultural sector. After a disaster in Region A, the lost wheat production cannot be substituted by rice from Region B. This information (if known) can be incorporated into the model by setting the corresponding trade flexibility factors to zero, preventing any trade between Region A and Region B following the disaster.

### 2.3 Total economic impact of a disaster

Disasters disrupt the economy through both production-side and consumer-side effects. From a production perspective, disaster-affected regions experience a decline in production, while other regions may exhibit either an increase or decrease in production. Moreover, in the aftermath of a disaster, industries may implement inefficient production processes, producing by-products (wasteful production) that are not utilized in the economy. On the consumer side, supply constraints may result in the rationing of final consumer products. Conventionally, economic models for disaster applications have predominantly focused on the production side of impacts, often neglecting the consumer side of impacts associated with supply shortages. For instance, in conventional input-output (I-O) models, economic losses are typically measured as the reduction in total output (Rose et al., 1997) or value-added (Hallegatte, 2014) when comparing pre- and post-disaster conditions. Koks &

Thissen (2016) extended this by defining disaster costs as the combined effect of changes in value-added and production inefficiencies, redistributed across regions using a cost-push I-O price model (Miller & Blair, 2022). However, consumer side impacts weren't considered.

In our model, we adopt a comprehensive approach by incorporating the following three components to estimate the total economic impacts of a disaster: (1) value of reduced supply of products (adjusted for rationing) (2) value of inefficiently produced products and (3) production equivalent value of rationed products (see Step 3, Table 1). The conversion of consumer side impacts (see Step 3; Table 1) into its production equivalent ensures that all the components of disaster impacts are on a common scale. While the component (1) can either be positive or negative, component (2) and (3) are always positive. We derive the total impacts from the 'lost demand for products (encompassing both inter-sector and final demand) between pre-and post-disaster conditions'.

Mathematically, the total impact $c$ is written as,

$$c = \sum_r \sum_p \left( d_{r,p}^{\,i} - d_{r,p} \right) \tag{1}$$

In pre-disaster conditions (represented by the superscript $i$), demand equals supply. However, post-disaster demand i.e., the value of efficient production, is calculated as the difference between supply and the quantity of inefficiently produced products $k$, as shown in equation (2).

$$d_{r,p} = s_{r,p} - k_{r,p} \tag{2}$$

Substituting these conditions in (1), we get

$$c = \sum_r \sum_p \left( s_{r,p}^{\,i} - s_{r,p} \right) + \sum_r \sum_p k_{r,p} \tag{3}$$

Next, the value of rationed products is segregated from the supply (or output) reduction and its production equivalent ($x'_{r,s}$) is added. The total impact is shown in equation (4).

$$c = \sum_r \sum_p \left( s_{r,p}^{\,i} - s_{r,p} - \bar{v}_{r,p} \right) + \sum_r \sum_p k_{r,p} + \sum_r \sum_s x'_{r,s} \tag{4}$$

## 2.4 Analysis for disaster risk applications

The proposed MRIA is capable of a comprehensive range of analyses of the economic consequences of disasters. Additionally, we show the importance of sensitivity, criticality, and incremental disruption analysis concerning a specific case using this model.

### 2.4.1 Sensitivity analysis

In this analysis, we examine the impacts of multi-sector disruptions within a single region, using the Dutch NUTS2 region Zuid-Holland as a case study. We simulate a scenario where all manufacturing sectors in the region experience a 10% disruption, similar to the effects of natural hazards like floods. The macroeconomic impacts (e.g., output changes and disaster trade) are evaluated under 18 different model setups, combining six production extension factors (0%, 1%, 2.5%, 5%, 7.5%, and 10%) and three trade flexibility factors (0%, 25%, and 100%) (see Constraints; Step1; Table 1). To conduct a more detailed analysis, we examine three specific systems characterized by production extension (and trade flexibility factors) of 0 (0) %, 1 (25) % , and 2.5 (100) %. These systems will be referred to as "rigid," "moderate," and "flexible," respectively, throughout the paper. These systems are analysed to understand the regional and sectoral-level interactions post-disruption.

### 2.4.2 Criticality analysis

In this criticality analysis, we identify the vulnerable (i.e., less redundant) sectors in the Dutch economy. Yu et al. (2014) defines critical sectors using various measures, including inoperability-based economic impact, sector output, and average propagation length (Dietzenbacher et al., 2005). In our analysis, we stress all the economic sectors (55 sectors x 12 regions) in the Netherlands by a disruption of 10% under a flexible system (i.e., with 2.5% production extension capacity and 100% trade flexibility). We define criticality based on the value of products rationed when stressed. Since the model prioritizes imports, higher levels of rationing indicate a lack of redundancy (i.e., lack of possible substitutes) in the economy, making these sectors more critical. Finally, we normalize the rationing values as shown in Equation 5. The numerator represents the total value of products rationed when the sector $s$ in the region $r$ is stressed

$$c_{r,s} = \frac{\left(\sum_r \sum_p \bar{v}_{r,p}\right)_{r,s}}{\sum_r \sum_s \left(\sum_r \sum_p \bar{v}_{r,p}\right)_{r,s}} \quad (5)$$

The proposed criticality measure is studied along with a normalised score based on economic output (follows equation 5 with rationing replaced by $\bar{x}_{r,s}$) and the location quotient (Miller et al., 1997) of sectors. The location quotient is a measure to identify the sectoral specialisation of a region as the relative concentration of economic activities in a region compared to an encompassing reference region. The location quotient of sector $s$ in region $r$ is given by Equation 6, where $x_{r,s}$ and $x_{n,s}$ represents the output of sector $s$ in region $r$ and the reference region $n$ (the Netherlands in this case) respectively, while $x_r$ and $x_n$ represent the total output of regions $r$ and $n$, respectively.

$$q_{r,s} = \frac{\left(\dfrac{x_{r,s}}{x_r}\right)}{\left(\dfrac{x_{n,s}}{x_n}\right)} \quad (6)$$

### 2.4.3 Incremental disruption analysis

The goal of the analysis is to identify critical transition points (e.g., the disruption threshold after which rationing of products is inevitable) with increasing disruptions and also to examine how the system's flexibility influences the shift in these critical points. For this analysis, we focus on two select sectors from Zuid-Holland,

Oil and Petrochemicals and the Chemicals sector. A series of disruptions ranging from 1% to 100% (with 1,2,5, 10 % disruptions, and in steps of 10% until 100%) is applied to these sectors with rigid and flexible parameters, i.e., with production extension (trade flexibility) of 0%(0) and 2.5% (100), respectively.

## 3. Results

### 3.1 Sensitivity analysis

As outlined in the Methods, the first analysis investigates the macroeconomic effects of a uniform 10% disruption across all manufacturing sectors in Zuid-Holland under varying parameters. Figure 1 illustrates the responses. Overall, increased system flexibility reduces impacts, including less rationed products, reduced change in output before and after the disaster. Also, the value of inefficiently produced products and disaster-related trade increases in the flexible regime.

Figure 1a presents the value of rationed products with increasing system flexibility. Systems with 0% trade flexibility exhibit distinct behaviour compared to others. In these systems, the amount of rationed products and disaster trade (see Figure 1e) remain insensitive to the changes in production capacities. For instance, increasing the production capacity from 0% to 2.5% results in only a 4.5 % reduction in rationed products from 19.3 to 18.4 million euros per day). In contrast, systems with 25% and 100% trade flexibility that allow for more regions to alleviate rationing by selling their products in the disaster region show reductions of 34% (from 15.4 to 10.2 million euros per day) and 49% (15.4 to 7.9 million euros per day), respectively.

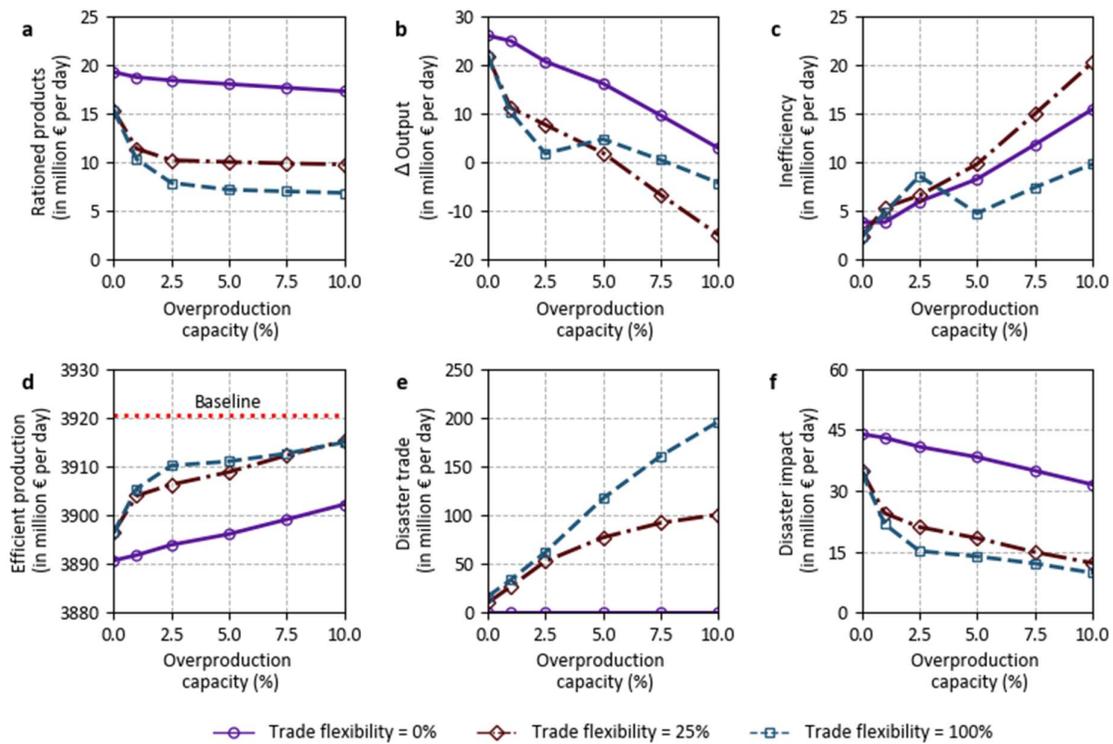

**Figure 1**: Outcomes of 10% disruption to manufacturing sectors in Zuid-Holland: (a) value of rationed products, (b) output change b/w pre and post-disaster, (c) inefficiency, (d) efficient production (see equation 2), (e) disaster trade, and (e) disaster cost (see equation 4).

The change in total output between pre- and post-disaster conditions decreases as production capacity increases (see Figure 1b). Eventually, it also enters the negative regime, i.e., post-disaster output exceeds pre-disaster levels in a few systems. However, this must be considered alongside wasteful (or) inefficient production, presented in Figure 1c. The value of inefficiently produced products increases with increased production capacity. For example, in a system with a 10% production extension and 25% trade flexibility, the post-disaster output is 14.86 million euros per day higher than pre-disaster levels. However, this comes at the cost of 20.3 million euros per day in wasteful production—products that are not utilized elsewhere in the economy. As a result, in Figure 1d, the efficient production (see equation 2) remains lower than the baseline production. This also explains the transition in system responses at and after 2.5% production extension limits with 100% trade flexibility. The more constrained 2.5% system had to meet demands through wasteful production, whereas the 5% system, with greater flexibility, was able to satisfy demands with relatively less inefficiency. Hence, the efficient production of the 5% system surpasses that of the 2.5% system.

Figures 1e and 1f present the value of disaster trade and the disaster impact (see equation 4) with increasing production capacity, respectively. It is observed that the divergence in responses between systems with 25% and 100% trade flexibility becomes increasingly evident as the production capacity limit rises, i.e., at lower production capacity, they exhibit similar responses. Higher production capacity leads to greater availability of products for disaster trade and while trade flexibility leads to better allocative distribution of the additional production over the different regions and allows for the most productive regions to take over production. These findings suggest that increased production capacity and flexible trade mechanisms must work in tandem to mitigate impacts effectively. Additional production can only be utilized efficiently if post-disaster trade channels are available. Otherwise, the economy will meet demands through inefficient production, which is obscured within the post-disaster economic output.

Further examination of situations with 25% and 100% trade flexibility reveals that the rate of impact reduction (e.g., measured by the value of reduced disaster impact) declines as production capacity increases, as it allows for better allocation of production over the regions. For instance, in the 100% flexible system (see Figure 1f), extending the production capacity of non-disrupted sectors by 2.5% decreases the value of the disaster cost by 19.3 (from 34.5 to 15.1) million euros per day. However, further expansion of production capacity yields only marginal benefits. This observation also holds for the value of rationed products (see Figure 1a). In general, sectors (or firms) have limited capacity to scale up production due to existing resource constraints, such as capital and labour. However, our results show that even a modest expansion in production can significantly reduce disaster impacts, especially if trade flexibility allows for regional allocative efficiency.

In Figure 2, we illustrate the regional variability of the effects using absolute changes in regional output across the 12 NUTS-2 regions under three system configurations: rigid, moderate, and flexible, as defined in Methods (see Section 2.4.1). As expected, under a rigid configuration, all regions experience a decline in output. Notably, Zuid-Holland, Noord-Holland, and Noord-Brabant exhibit daily output reductions of 20.1, 3.0 million, and 1.4 million euros per day, respectively. This response aligns with the outcomes of a traditional Input-Output (IO) model, where substitution of products between regions is not allowed. In contrast, regions exhibit positive effects under moderate and flexible configurations. For example, under the flexible system, Utrecht and Gelderland experience notable output reductions of 15.1 million and 2.7 million euros per day, respectively, following Zuid-Holland. Conversely, Noord Holland and Noord Brabant display positive effects, with daily output gains of 16.8 million and 4.6 million euros per day, respectively. This is because Noord Holland and Noord Brabant are the key economic regions of the Dutch economy after Zuid-Holland. Their larger economic capacities allow them to produce more products and allocate efficiently to other regions to compensate for the lost production in Zuid-Holland. For example, a fraction of the products initially produced in Utrecht are now made by these gaining regions, while Utrecht meets its demands through imports. As a result, although Utrecht experienced a production decline, they did not face shortages in final consumption products, i.e., no rationing, as imports fulfilled their needs. This pattern is also reflected in the post-disaster trade heatmap presented in Supplementary Figure 3(a-c), which shows Zuid-Holland as the largest recipient of disaster imports, followed by Utrecht and Gelderland.

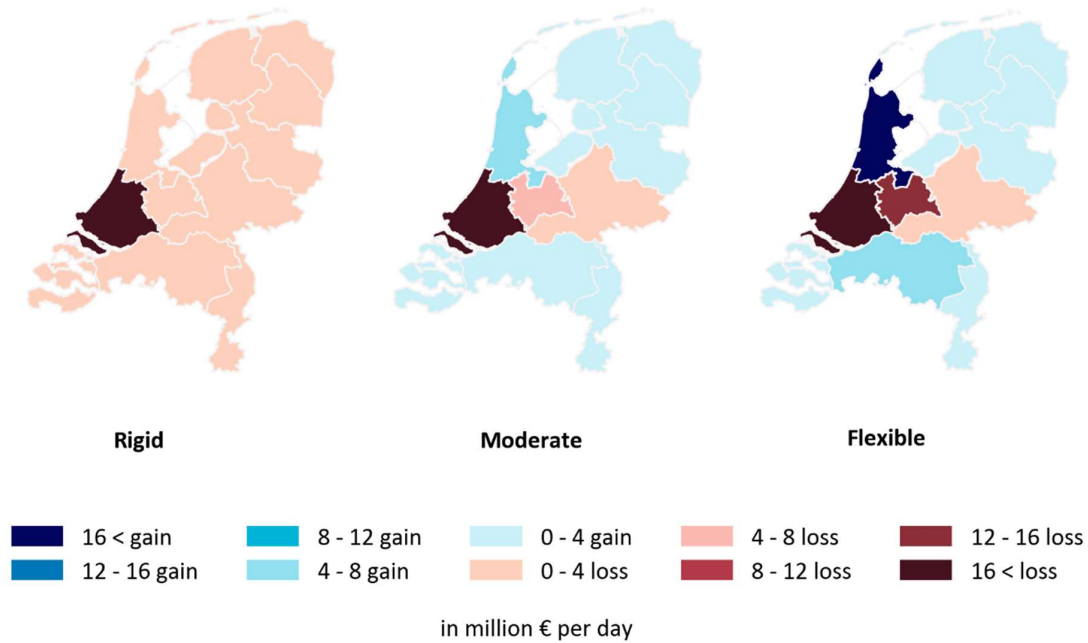

**Figure 2**: Regional absolute output changes under (a) rigid, (b) moderate, and (c) flexible systems

Further, we also observe that the differing regional effects are also driven by the region's dependency on the disrupted sectors (manufacturing sectors of Zuid Holland in this case). Among the four regions discussed above, Noord Holland and Noord Brabant are predominantly export-oriented, supplying more products to Zuid-Holland's manufacturing sectors than they consume. In contrast, Utrecht and Gelderland are import-oriented, receiving more products from Zuid-Holland's manufacturing sectors than they supply. To quantify these trade dependencies, we calculated the ratio of this supply to consumption for each region. The results, presented in Supplementary Table 3, reveal a ratio of 1.87 for Utrecht, indicating that for every unit of products consumed by Zuid-Holland's manufacturing sectors from Utrecht, 1.87 units are supplied back to it. Meanwhile, Noord Holland and Noord Brabant exhibit ratios below 1, highlighting their export-oriented relationship with Zuid-Holland's manufacturing sectors.

Sectoral-level impacts further illustrate the model behaviour and its response to disruptions. For instance, consider the chemicals sector (C20) of the Netherlands. The major producers are Zuid-Holland, Noord Brabant, and Limburg, with outputs of 30.5, 23.1, and 21.7 million euros per day, respectively, while the rest of the regions produce relatively minimal amounts. Following the disruption in Zuid-Holland, under a flexible system, the chemical sectors in Noord Brabant and Limburg display positive effects (see Supplementary Figure 4a), reflecting the extension of production to compensate for the lost output, which results in increased imports of chemical products from Noord Brabant and Limburg. This observation confirms that the model operates as intended, identifying and utilizing alternative suppliers to mitigate the impacts of disruptions.

But, the analysis thus far assumes product homogeneity, meaning that all regions within a given product category produce identical and substitutable products. However, this is often not the case. As a counterfactual scenario, we consider a case where Noord Brabant produces a distinct set of chemical products that differ entirely from those produced in the rest of the Netherlands. Under this assumption, the lost production in Zuid-Holland cannot be compensated by Noord Brabant. This scenario is incorporated into the model by adjusting the trade flexibility factor, setting all disaster-related trade of chemical products between Noord Brabant and other regions to zero. The results, presented in Supplementary Figure 4b, demonstrate that under these constraints, the chemical sector of Noord Brabant does not exhibit positive effects, highlighting the model's adaptability. Also, expanding the country-level model to include additional European regions is expected to mitigate the observed impacts. A larger set of alternative producers in neighbouring countries would enhance the system's ability to compensate for production losses. For example, disruptions to refineries near the Port of Rotterdam could be offset by refineries near the Port of Antwerp in Belgium.

### 3.2    Criticality analysis

This analysis identifies critical sectors in the Dutch economy by stress-testing individual sectors (a disruption of 10% applied to the sector individually with flexible parameter settings) and assessing the resulting value of rationed products. Rationing the final consumption occurs when alternate suppliers are unavailable or imports and production are constrained, revealing vulnerabilities. Sectors leading to higher levels of rationed products are considered 'critical' due to insufficient system redundancy. Fig 3a. presents the normalized criticality score based on rationing (see Equation 5). Zuid-Holland's Coke and Refined Petroleum sector (C19) emerges as the most critical due to its reliance on Rotterdam's oil refineries, which produce over 95% of the Netherlands' total output. This regional concentration makes the system highly vulnerable to disruptions, with significant ripple effects across the economy. Similarly, Noord Holland's Air Transport (H51) and Financial Services (K) sectors are also identified as critical. Again, the criticality of air transport services in Noord Holland can be justified because of Schiphol, one of Europe's largest transfer hubs.

The outcomes of the analysis are further interpreted using the economic output and location quotient (LQ) of sectors. A comparison of Figures 3a and 3b reveals that sectors identified as critical tend to exhibit relatively high levels of economic output. For instance, the Coke and Refined Petroleum sector (C19) in Zuid-Holland, which is classified as critical in our analysis, is also the second-largest producing sector in the Netherlands. However, the converse does not always hold true. For example, the Food and Beverages sector (C10T12) is one of the Netherlands' largest economic contributors. However, a substantial proportion of the C10T12 sector exists across all regions (i.e., geographically dispersed production), making substitution from other regions feasible following a disaster. A similar pattern is observed for the Chemical (C20) and the Construction (F) sectors.

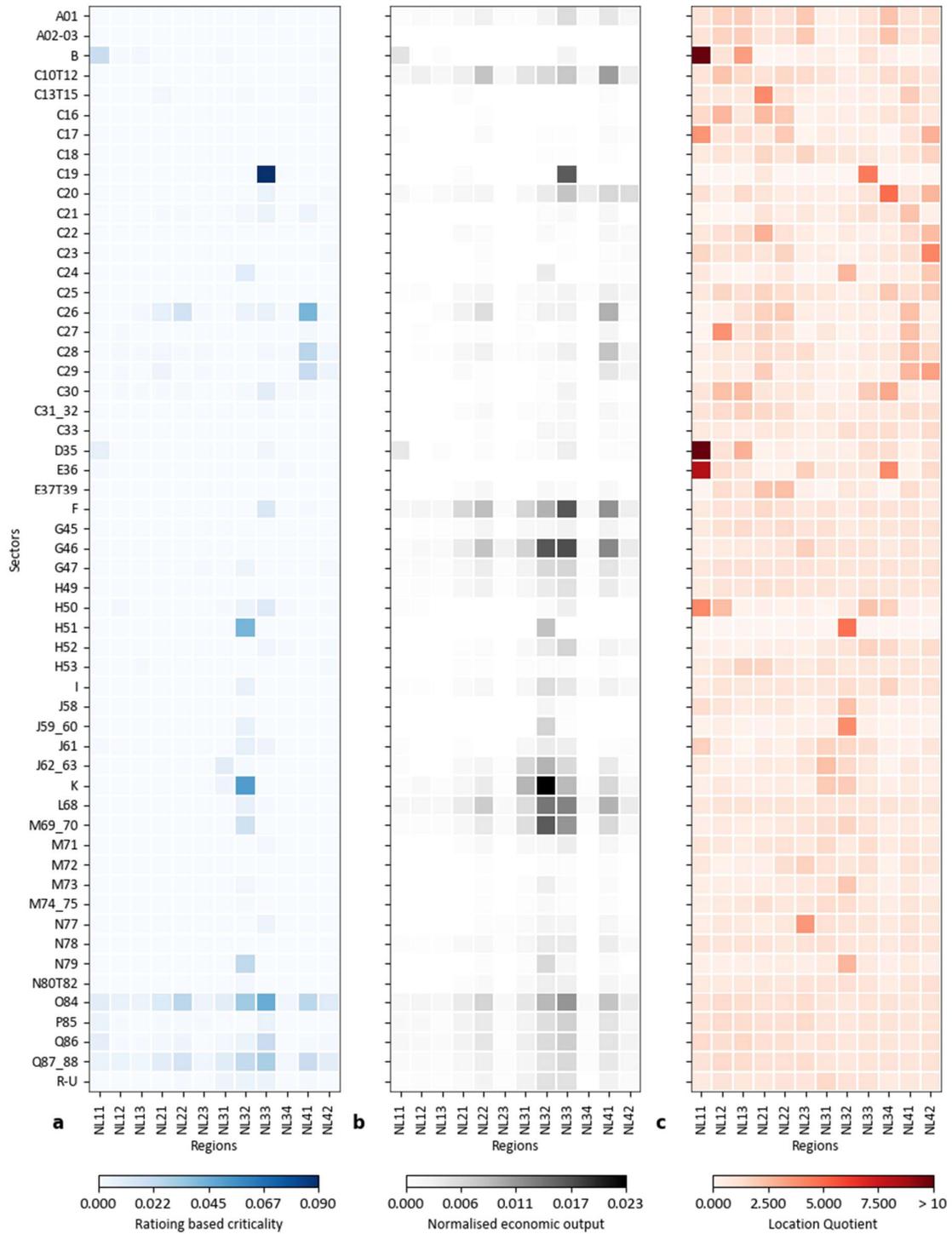

**Figure 3**: Criticality of sectors based on (a) the value of rationed products (under 10% disruption – flexible system) when stressed, (b) the economic output, and (c) the location quotient of the sector (Refer to Supplementary Table 1 for the description of sectors).

On the other hand, the LQ (see Figure 3c) measures the relative importance of a sector within a region (the numerator of Equation 6) against its importance at the national level (the denominator of Equation 6). For instance, Zuid-Holland's Coke and Refined Petroleum sector (C19) and Noord Holland's Air Transport sector

(H51) exhibit location quotients exceeding 4.5, indicating their relatively high regional economic importance, while the C10T12 sector has an average LQ of 1.2 across all regions. Additionally, the Mining and Quarrying sector (B) in Groningen has the highest location quotient (16.7) in the national economy, and also attains a relatively high rationing score (0.02 – 97.5th percentile) in our analysis. These findings validate the efficacy of the proposed approach in identifying critical sectors not solely based on their economic output, but also by considering the potential for regional substitution within the economy following disasters.

### 3.3  Incremental disruption analysis

Incremental disruption analysis, also known as transition analysis, is a useful approach for identifying critical disruption thresholds that define the system's resilience. These thresholds for a sector are influenced by key factors, including the availability of alternative suppliers, the system's adaptability to disruptions, and the sector's maximum capacity to ration its products. For this study, we focus on two sectors in Zuid-Holland: (a) Chemicals (C20), which represents a sector with access to alternative suppliers, and (b) Coke and Refined Petroleum (C19), identified as the most critical sector (see Section 3.2).

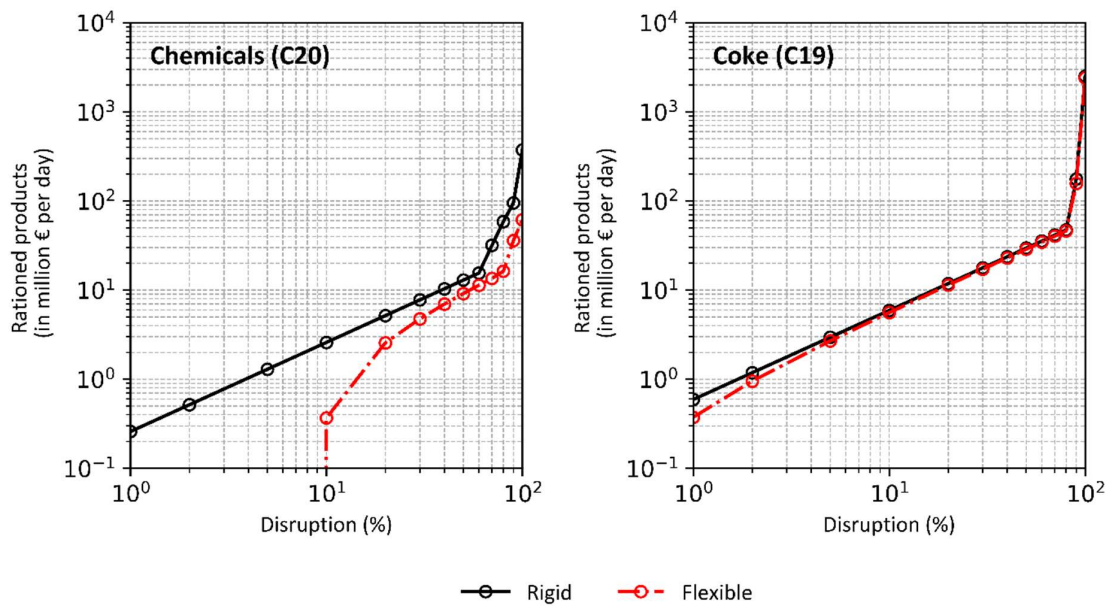

**Figure 4:** Incremental disruption analysis of Zuid Holland's chemical (C20) and coke and refined petroleum (C19) sector.

Figure 4a illustrates the value of rationed products with increasing disruption to the chemical sector in Zuid-Holland, comparing outcomes under rigid and flexible parameter settings. For a flexible system, our analysis reveals that the value of rationed products transitions through three distinct phases as disruption increases: i) the 'zone of no rationing,' ii) the 'zone of limited rationing,' and iii) the 'zone of rationing cascade'. There are no products rationed until a 10% disruption (marking the 'zone of no rationing'). This is because imports increase steadily up to a 10% disruption and saturate thereafter (see Supplementary Figure 5). Beyond this point, rationing gradually increases until it reaches the maximum possible value (i.e., the sum of final consumption and exports of chemical products from Zuid Holland) at 80% disruption. After this threshold, the total value of rationed products escalates sharply due to cascading effects, as products begin to be greatly rationed in other regions as well. With almost no imports, the rigid system is forced to ration from the onset of minimal disruptions. Also, the zone of the rationing cascade arrives much earlier at a 60% disruption threshold compared to the flexible system. Similarly, Figure 4b presents the value of rationed products with increasing disruption to the coke and refined petroleum sector. Notably, there is no distinction between the rigid and flexible systems due to the absence of alternative suppliers in the economy, eliminating the 'zone of no rationing.'

4. **Discussion**

We confirm Rose's (2004) assertion that "*point estimates exaggerate the certainty of the analysis*," emphasizing that presenting single estimates of macroeconomic losses can overstate the reliability of the results. Often, macroeconomic models designed for disaster risk applications differ in their modelling assumptions and parameters which can greatly affect their outcomes. Consequently, decisions based on single-point estimates could be prone to a high margin of error. In this article, we illustrate this by showing how responses to the same set of disruptions can differ substantially under varying production extensions and trade flexibilities.

A key distinction of our model from existing macroeconomic impact models lies in its integrated approach to handling post-disaster production limits and trade constraints, for instance due to logistical limitations. Our rigid model closely replicates the outcomes of traditional input-output models (Okuyama, 2004; Rose et al., 1997) when rationing does not occur. On the other hand, ARIO models, which incorporate production extension but not logistical constraints, often exhibit a nonlinear relationship between direct production losses and indirect cascading effects (Hallegatte, 2008, 2014). Unlike MRIA, ARIO does not account for the interregional substitution of products, thereby overlooking the benefits of support from neighboring regions. Surprisingly, the multi-regional ARIO model used in Liu et al. (2023) also demonstrates such cascading effects, largely due to a very low production extension factor (0.3%) combined with severe disruptions. Cascading effects are also observed in MRIA under extreme disruptions (refer to Figure 4a), where the sectors reach their maximum possible production and rationing limits, leading to a zone of negative ripple effects. Next, in comparing MRIA with the optimization model developed by Oosterhaven & Bouwmeester (2016), the primary distinction lies in their objectives. MRIA prioritizes minimizing rationing by enabling producers to meet demand at the lowest cost, whereas their model focuses on maintaining proximity to the pre-disaster equilibrium, potentially leading to different outcomes. Furthermore, the absence of production extension and logistical constraints in their model makes the results of MRIA more interpretable.

MRIA also shares similarities with agent-based models (ABM) built for disaster risk applications (Bierkandt et al., 2014; Otto et al., 2017; Wenz et al., 2014; Willner et al., 2018). For example, both MRIA and ABM models incorporate production extension and logistical constraints. However, ABMs employ a more sophisticated modelling framework by explicitly considering logistical constraints in the actual transportation of products, taking geographical distance into account. Additionally, ABMs are dynamic and incorporate storage capacity (represented by the number of buffer days available to sustain production in the absence of inputs), a feature not present in MRIA but one that could be integrated in future developments. This absence of buffer capacity might make MRIA estimate relatively higher impacts compared to ABMs. Furthermore, ABMs do not explicitly formulate the assumption of minimizing rationed products, which is reflected in their results (Bierkandt et al., 2014), where final demand sites experience greater impacts. Notably, the pattern of diminishing returns with increasing production extension factors (see Figure 1a and 1f; refer to Section 3.1) is also observed in Wenz et al. (2014), reinforcing the consistency of macroeconomic insights across different modelling approaches.

Our model suggests that disaster impact reduction is most effective when both post-disaster production limits and logistics flexibility are increased together. This indicates that policy decisions focused only on enhancing production resilience while ignoring logistics resilience will not yield optimal benefits, and vice versa. For instance, the same level of disaster risk reduction can be achieved by increasing 1% production and 100% trade flexibility, or by increasing 2.5% production and 25% trade flexibility (see Figures 1a and 1f). However, the costs associated with expanding these capacities may differ. This allows for the identification of optimal systemic strategies to minimize disaster risk. Secondly, we find that unaffected larger economic regions with export-focused relationships towards the disrupted region (and sectors) display positive effects during disasters. Strengthening the resilience of these regions is crucial, as their ability to rapidly scale production and maintain supply chain stability enhances overall economic recovery. Also, we employ a straightforward criticality analysis to identify key sectors, which is not only valuable for policymakers but also for stakeholders (e.g., infrastructure owners) seeking optimized strategies for climate adaptation. For example, consider a transmission network provider with limited financial resources may need to prioritize the reinforcement of certain elements within their network. Such a macroscopic view of criticality enables the selection of network elements that support the most critical sectors, e.g., the substation and transmission elements powering the oil refineries in Zuid-Holland. While local-level prioritization may reduce damages, system-level criticality-based prioritization could significantly reduce the cascading effects. Additionally, the transition analysis (see Section 3.3) conducted with our model helps identify the required inventory capacity for sectors under varied disruption levels. Our model effectively supports these applications.

Future enhancements of the model will focus on incorporating dynamic effects through a fully time-dependent framework. Macroeconomic impacts are time-dependent, typically peaking immediately after a disaster and subsiding as recovery efforts take hold. Currently, our model solves one time step at a time, i.e., recursive dynamic, without linking the outcomes of previous time steps to those of future ones. However, adaptative responses like expansion of the production capacity are usually dynamic processes that improve over time (Hallegatte, 2008) to counteract disaster-induced disruptions. Inclusion of such effects results in a more realistic representation of post-disaster dynamics and will be a key direction for future model development.

## 5. Conclusion

In this paper, we presented a Multi-Regional Impact Assessment (MRIA) model, which builds upon the foundations established by Koks & Thissen (2016). MRIA is a multi-sector, supply-and-use macroeconomic model that utilizes linear optimization techniques to assess the macroeconomic impacts of disasters. When disruptions occur due to hazards, MRIA enables alternative suppliers to increase their production capacity and meet demands through additional post-disaster trade. The model also applies demand rationing to allocate resources efficiently if no alternative suppliers are available. This model improves upon Koks & Thissen (2016) in two key ways: (a) by capturing regional-level post-disaster trade patterns and (b) by explicitly incorporating logistical constraints through a trade flexibility factor in post-disaster scenarios.

The model was first applied to assess the macroeconomic effects of manufacturing disruptions in Zuid-Holland, Netherlands, considering different levels of production extension and trade flexibility. Our analysis highlights the importance of logistical constraints in the model, as logistically constrained systems (i.e., with limited trade flexibility after a disaster) experience greater impacts than systems with unrestricted trade flexibility. Moreover, our analysis emphasizes the necessity of increasing production capacity and trade flexibility in tandem. Expanding production capacity alone, without a corresponding increase in trade flexibility, results in suboptimal outcomes. Additionally, the model formulation effectively captures regional trade patterns following a disaster while also allowing for the integration of trade restrictions if required. For example, it enables the relaxation of the assumption of product homogeneity for substitution between regions. At the regional level, we observe significant negative impacts in the disrupted region. While larger, export-oriented economic regions experience positive effects primarily due to their ability to expand production capacity and meet the economy's demand.

Subsequently, we proposed a criticality approach to identify key sectors within the economy, as demonstrated in the Dutch case study. This approach employs unmet final demand as a measure for determining sectoral criticality. Its novelty lies in its ability to account for both the flexibility of the economic system and the magnitude of disruptions while capturing the potential for substitution possibilities between regions in the post-disaster scenario. Finally, we conduct an incremental disruption analysis for select sectors in Zuid-Holland. Our findings identified three distinct response phases as disruption levels increased. Initially, sectors manage disruptions with support from other regions ("zone of no rationing"). As disruptions intensified, sectors began rationing products confined to the sector's possible rationing limits ("zone of limited rationing"). Under extreme disruption, rationing extends to products from other regions and sectors ("zone of rationing cascade").

The results presented in this paper are at the national scale; however, the model can be readily scaled up to a Pan-European level. Future enhancements to the model will incorporate inventory capacities and dynamic effects to capture impacts over time. We believe that the MRIA model will serve as an effective tool for assessing disaster impacts and informing the development of disaster risk mitigation strategies at both a regional and a macroeconomic scale.

**Disclosure statement:**

No potential conflict of interest was reported by the authors. The authors acknowledge the use of OpenAI's ChatGPT for text editing and refinement of the manuscript. No content was generated by the tool without authors' oversight, and all scientific interpretations, analyses, and conclusions are the authors' own.

**Author contributions:**

SR: Conceptualization, Methodology, Analysis, Post-processing, Writing – original draft, review, and editing. MT: Conceptualization, Methodology, Writing –review and editing, and Supervision. MR: Writing –review and editing, and Supervision. EK: Conceptualization, Methodology, Writing –review and editing, Supervision, and Funding Acquisition.

**Funding:**

This work was funded by EU's HORIZON -MISS-2021-CLIMA-02 under the grant agreement No: 101093854 for the project titled 'Multi-hazard Infrastructure Risk Assessment for Climate Adaptation – MIRACA'. MR received support from the MYRIAD-EU project, which received funding from the European Union's Horizon 2020 research and innovation programme under grant agreement No. 101003276.. EK received funding from the Dutch Research Council (NOW) under grant agreement No. VI.Veni.194.033.

**Data availability:**

The codes and input data used in this project are openly accessible and can be found at https://doi.org/10.5281/zenodo.15119803. This repository includes the full set of scripts and input files necessary to reproduce the analyses and figures. The basemap used in the Figure 2 is downloaded from https://service.pdok.nl/cbs/gebiedsindelingen/atom/v1_0/gebiedsindelingen.xml (last accessed 16-04-2025).

**Supplementary Table 1:** List of sectors

| Sectors | Product / Sector description |
|---|---|
| A01 | Products of agriculture, hunting and related services |
| A02-03 | Products of forestry, logging and related services, fishing and aquaculture |
| B | Mining and quarrying |
| C10T12 | Food, beverages and tobacco products |
| C13T15 | Textiles, wearing apparel, leather and related products |
| C16 | Wood and of products of wood and cork, except furniture; articles of straw and plaiting materials |
| C17 | Paper and paper products |
| C18 | Printing and recording services |
| C19 | Coke and refined petroleum products |
| C20 | Chemicals and chemical products |
| C21 | Basic pharmaceutical products and pharmaceutical preparations |
| C22 | Rubber and plastic products |
| C23 | Other non-metallic mineral products |
| C24 | Basic metals |
| C25 | Fabricated metal products, except machinery and equipment |
| C26 | Computer, electronic and optical products |
| C27 | Electrical equipment |
| C28 | Machinery and equipment n.e.c. |
| C29 | Motor vehicles, trailers and semi-trailers |
| C30 | Other transport equipment |
| C31_32 | Furniture and other manufactured goods |
| C33 | Repair and installation services of machinery and equipment |
| D35 | Electricity, gas, steam and air conditioning |
| E36 | Natural water; water treatment and supply services |
| E37T39 | Waste and sewage management services |
| F | Constructions and construction works |
| G45 | Wholesale and retail trade and repair services of motor vehicles and motorcycles |
| G46 | Wholesale trade services, except of motor vehicles and motorcycles |
| G47 | Retail trade services, except of motor vehicles and motorcycles |
| H49 | Land transport services and transport services via pipelines |
| H50 | Water transport services |
| H51 | Air transport services |
| H52 | Warehousing and support services for transportation |
| H53 | Postal and courier services |
| I | Accommodation and food services |
| J58 | Publishing services |
| J59_60 | Motion picture, video and television programme production services, sound recording and music publishing; programming |
| J61 | Telecommunications services |
| J62_63 | Computer programming, consultancy and related services; Information services |
| K | Financial services, except insurance and pension funding, Insurance, reinsurance and pension funding services, except |
| L68 | Real estate services |
| M69_70 | Legal and accounting services; services of head offices; management consultancy services |
| M71 | Architectural and engineering services; technical testing and analysis services |
| M72 | Scientific research and development services |
| M73 | Advertising and market research services |
| M74_75 | Other professional, scientific and technical services and veterinary services |
| N77 | Rental and leasing services |
| N78 | Employment services |
| N79 | Travel agency, tour operator and other reservation services and related services |
| N80T82 | Security and investigation services; services to buildings and landscape; office administrative, office support and other |
| O84 | Public administration and defence services- compulsory social security services |
| P85 | Education services |
| Q86 | Human health services |
| Q87_88 | Residential care services; social work services without accommodation |
| R-U | Creative, arts, entertainment, library, archive, museum, other cultural services; gambling and betting services, Sporting |

**Supplementary Table 2:** MRIA-v1 variables and parameters

| Symbols | Description |
|---|---|
| | **Endogenous variables** |
| $x_{r,s}$ | Output of the sector $s$ in the region $r$ after the disaster |
| $v_{r,p}$ | Rationing of a product $p$ from the region $r$ after the disaster |
| $t_{r',r,p}$ | Disaster trade of the product $p$ from region $r'$ to region $r$ |
| $s_{r,p}$ | Total supply of a product $p$ from the region $r$ |
| $d_{r,p}$ | Total demand for a product $p$ from the region $r$ |
| | **Exogenous variables** |
| $\bar{f}_{r,p}$ | Final consumption of a product $p$ from the region $r$ |
| $\bar{e}_{r,p}$ | Export of product $p$ from region $r$ to the rest of the world (ROW) |
| $\bar{n}_{r,p}$ | Change in demand for a product $p$ from the region $r$ if any after the disaster |
| $\bar{x}_{r,s}$ | Output of the sector $s$ in the region $r$ before the disaster |
| | **Model coefficients** |
| $C_{r,s,p}$ | Supply coefficient of the product $p$ by sector $s$ in the region $r$ |
| $B_{r,p,r',s}$ | Use coefficient of product $p$ from the region $r$ by sector $s$ in the region $r'$ |
| | **Model parameters** |
| $\delta_{r,s}$ | Production extension factor of sector $s$ in region $r$ |
| $\phi_{r',r,p}$ | Trade flexibility from region $r'$ to region $r$ for the product $p$ |
| $\alpha$ | Weighing parameter for disaster trade component in the objective function |
| | **Intermediates** |
| $\bar{v}_{r,p}$ | Rationing of a product $p$ from the region $r$ after the disaster after Step 1 |
| $\bar{\bar{v}}_{r,p}$ | Rationing of a product $p$ from the region $r$ after the disaster after Step 2 |
| $x'_{r,s}$ | Output required from the sector $s$ in the region $r$ to satisfy the demand vector $\bar{\bar{v}}_{r,p}$ |

**Supplementary Table 3:** Regional dependencies between manufacturing sectors of Zuid-Holland (NL 33) and other major economic regions of the Netherlands.

| NUTS-2 ID | Region | The total value of NL33 manufacturing products consumed (exported) by the region (in million Euros per day) (1) | The total value of the region's products consumed (imported) by manufacturing sectors of NL33 (in million Euros per day) (2) | Export / Import ratio (1) / (2) | Output of the region (in million Euros per day) |
|---|---|---|---|---|---|
| NL 32 | North Holland | 10,70 | 14,01 | **0,76** | 814,75 |
| NL 41 | North Brabant | 11,39 | 16,85 | **0,68** | 625,09 |
| NL 22 | Gelderland | 7,69 | 5,78 | **1,33** | 396,65 |
| NL 31 | Utrecht | 4,90 | 2,63 | **1,87** | 319,29 |

Note: A ratio less than '1' indicates that the region is export-focused towards the manufacturing sectors of Zuid-Holland and a ratio greater than '1' suggests vice versa. This ratio has to be interpreted in conjunction with the region's output because higher output increases the scope for overproduction and imports, and hence can potentially reflect more gains.

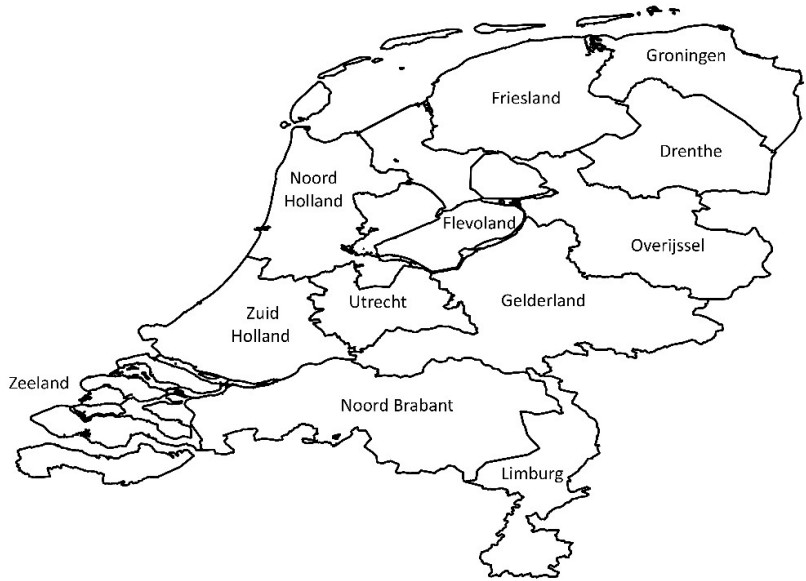

**Supplementary Figure 1**: NUTS-2 level administrative map of Netherlands

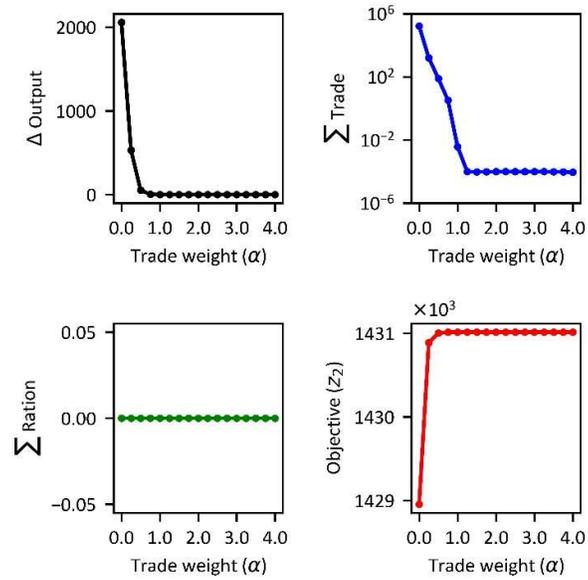

**Supplementary Figure 2**: Sensitivity of MRIA model variables (output, post-disaster trade, rationing, and the value of objective function) to the trade weight parameter ($\alpha$) under zero disruption.

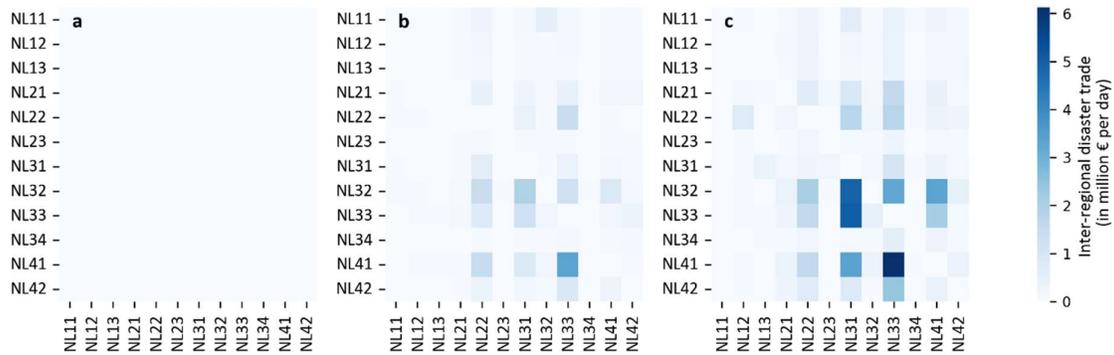

**Supplementary Figure 3**: Inter-regional disaster trade for (a) rigid, (b) moderate, and (c) flexible systems

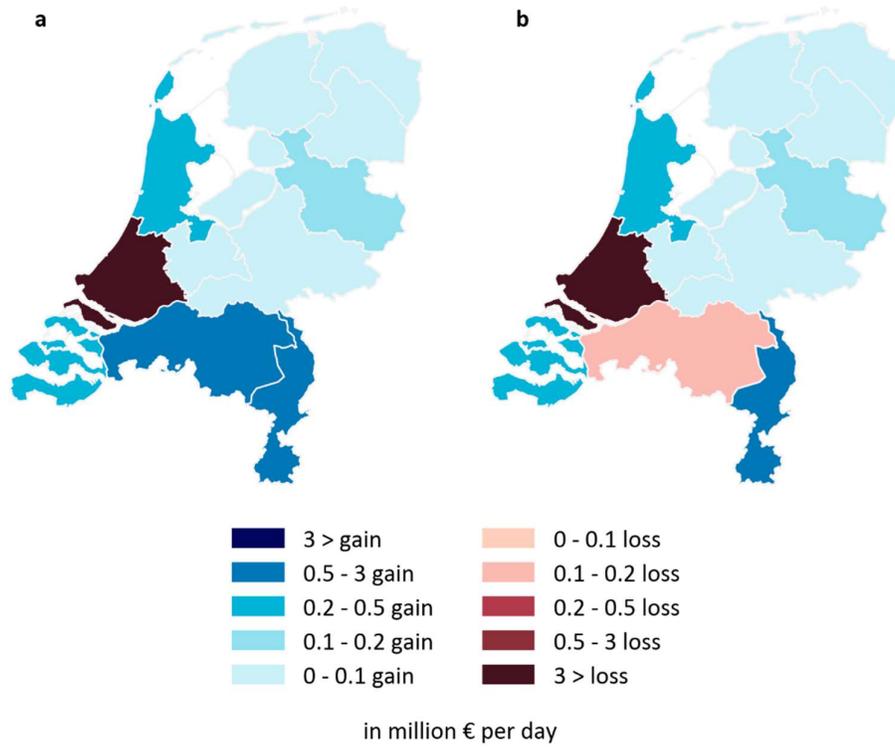

**Supplementary Figure 4**: Output change of chemical sector C20 under a flexible system (a) without and (b) with export restrictions applied to North Brabant ( a case of product heterogeneity)

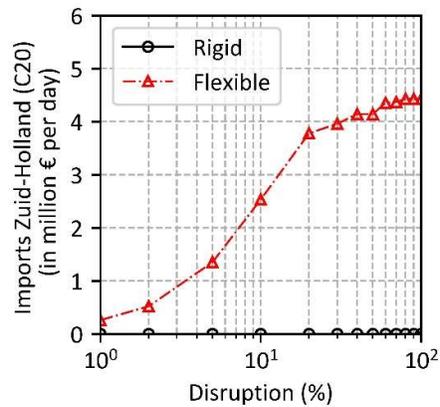

**Supplementary Figure 5**: Total import of products of chemical sector to Zuid-Holland with increasing disruption